\documentstyle [preprint,eqsecnum,aps]{revtex}

\begin{document}
\draft
\preprint{SNUTP-94-60, hep-th/9407048}

\title{
Abelian Chern-Simons field theory\\
and anyon equation on a sphere
}

\author {N. W. Park}

\address
{Department of Physics,
Chonnam National University,
Kwangju, 500-757, Korea}

\author {Chaiho Rim and D. S. Soh}
\address
{Department of Physics,
Chonbuk National University,
Chonju, 560-756, Korea}

\date {July 7, 1994}

\maketitle

\begin{abstract}

We analyze the Chern-Simons field
theory  coupled
to non-relativistic matter field  on a sphere
using canonical transformation on the fields
with special attention to the role of the rotation symmetry:
$SO(3)$ invariance
restricts the Hilbert space to the one
with a  definite number of charges
and dictates  Dirac quantization condition
to the Chern-Simons coefficient,
whereas $SO(2)$ invariance does not.
The corresponding
Schr\"odinger equation for many anyons (and for multispecies)
on the sphere are presented
with appropriate boundary condition.
In the presence of an external magnetic monopole source,
the ground state solutions
of anyons are compared with monopole harmonics.
The role of the translation and
modular symmetry on a torus is also expounded.

\end{abstract}


\narrowtext

\section{Introduction}

The statistical property of quantum mechanical particles is  known  in
two categories, fermion and boson. The dynamics of these particles  is
described systematically in second-quantized field  theories.  In  two
space dimension, however, anyonic statistics
\cite{leinaas} is  known  to  be
possible.  This  observation  has  come  not  only  from   theoretical
curiosity but also from experiment such  as  fractional  quantum  Hall
effect  \cite{girvin}.
One  may  even  go  further  to  define   fractional
statistics based on Hilbert space counting independent of space
dimension \cite{haldane}.

The dynamics of anyon
\onlinecite{arovas,goldhaber,semenoff,matsuyama,swanson}
is  described  in  the  first  quantized
theory  by  introducing statistical gauge potential minimally  coupled
to  the  originally bosonic or fermionic theory. After singular  gauge
transformation  (to  make  the  singular   gauge   transformation   is
meaningful,  one  has  to apply the  so-called  ``hard-core"  boundary
condition), the statistical gauge potential can  be  removed  and  the
wavefunction becomes multi-valued instead (statistical transmutation),
from which the name anyon stems. The second quantized theory of anyons
is yet to come \cite{nair}.
Nevetherless, One can relate Chern-Simons (CS)
field theory \onlinecite{deser}
in  two  space  and  one  time  dimension  to  the  first
quantized anyon theory:
One   can  obtain  the  corresponding
anyon  equation  using  the   second   quantized   field  theory   and
Heisenberg equation \cite{jackiw}.

The pure CS gauge action has no dynamical gauge field.
By coupling the  gauge   field
to an external source, one can investigate quantities such  as  Wilson
loops.  In  fact,   the   correlators   of   Wilson   lines   in   the
non-abelian theory are related with the polynomial invariant  of  knot
theory \cite{witten}.
When the gauge field is coupled to a dynamical  matter
field, it plays the additional role of statistical gauge potential  to
the matter field, which is best seen  in  the  abelian CS theory on an
infinite plane \onlinecite{semenoff,matsuyama,swanson}.
Its non-abelian statistical property is also investigated
using conformal blocks \cite{verlinde}.

In quantizing the CS theory, some subtleties  are  noted  due  to  the
Gauss constraint.
By solving the constraint first
and quantizing  the theory later,
one may lose some  information  inherent  to
the quantized theory such as quantum  holonomy  \cite{dunneja}.
In addition, restricting the Hilbert space
to the one with definite
particle number states
and treating the number operator  as a $c$-number
\cite{hosotani,fayya}
arouse controversies
\cite{hagsu}.
On torus, this procedure
makes the abelian CS coefficient rational
which depends on the number of particles:
The theory should pick up a certain
coefficient to accommodate the $N$ particles.
On the other hands,
one can quantize  the gauge field first
and restrict the Hilbert space later
according to the Gauss constraint
(not necessarily onto the definite
particle number states).
If one identifies
the fundamental domain of the torus with a
rectangular unit cell
and applies  the periodic (translation)  symmetry
in the both direction \cite{cr},
then one can treat the theory field-theoretically
and does not need the abelian CS coefficient  quantization.

The origin of the above discrepancy
(except the quantum holonomy)
may not be entirely due to the way of
gauge field quantization.
One can suspect that
this comes from the conflict with the
symmetry that the system is allowed to have
({\it e.g.} translation invariance
and/or modular invariance),
since
the Gauss constraint
(like in the presence of an external monopole source)
can not be realized arbitrarily in general
due to  the inherent symmetry of the compact manifold.
This paper is devoted to
illuminate the role of the symmetry of the
compact manifold (sphere and torus) on the
CS field theory and its connection to
the anyon equation.
For this purpose,
we start with  investigating CS theory on a sphere
field-theoretically following the method given
in \cite{cr},
which is simpler than the theory on a torus
since harmonic one form is absent.
(Related works can be found in \cite{lee,ienle}
where the Gauss constraint is treated
classically.)
The same field-theoretic analysis can be
used for the torus symmetry.

This paper is organized as follows.
In section II we present the mode decomposition of the gauge field  on
two patches of the sphere.
In section III  we  couple  the  non-relativistic  matter
field (fermion for definiteness)
to the gauge field.  The  hamiltonian
is re-expressed in terms of
canonically transformed new  field,  which
effectively solves the Gauss constraint.
The expression of the constraint and
hamiltonian is incorporated with
the rotation symmetry,
$SO(2)$ or $SO(3)$.
The Hilbert space restriction
due to the rotation invariance is
investigated in detail.
In  section  IV   we present anyon   equation
for $N$-identical particles
and for  multi-species  with the vanishing  total charge.
The boundary condition for the wavefunction
with the rotation symmetry taken care of.
In section V we  introduce  an  external  magetic
monopole source and discuss its ground state
and degeneracy.
It is noted that the ground state solution
of anyons is not annihilated by  the  ladder  operator  of
$su(2)$ unlike for the monopole harmonics.
Section VI  is  the  summary
and discussion, where we discuss about the
translation and modular symmetry on the torus
and its consequences.

\section{Two patch description of gauge field}

The gauge field on a sphere is a singular function
in general (connection on a $U(1)$-bundle).
Therefore, one needs coordinate patches
to describe the gauge field on
the whole space \cite{naka}.
We introduce
two coordinate patches, which are
two projected planes from the south and north poles
respectively (FIG.\ref{fig.1}).
The north (south) patch covers the whole points on the sphere
except the south (north) pole.
Let us denote a point as $\Omega=(\theta, \phi)$ on the sphere.
The same point is described as $P_N = \{\vec x_N\}$
on the north patch
($P_S = \{ \vec x _S\}$ on the south), where
\begin{eqnarray}
\vec x_N &=& (x_N^{(1)}, x_N^{(2)})
=(\tan {\theta \over 2} \cos \phi,
\tan {\theta \over 2} \sin \phi).
\nonumber\\
\vec x_S
&=& (\cot {\theta \over 2} \cos (\pi-\phi),
           \cot {\theta \over 2} \sin (\pi -\phi)).
\end{eqnarray}
The two representations have a relation
$
\vec x_S = (-x_N^{(1)}, x_N^{(2)})/ \vec x_N ^2
$,
which is in complex notation
$z_S = - 1 / z_N$ where $z= x^{(1)} + i x^{(2)}$.
The metric $g_{\mu \nu}$ in either patch is diagonal and is given as
\begin{equation}
g_{00} = 1,\qquad g_{ij} = - ({2 \over 1+ \vec x^2})^2 \delta_{ij}.
\end{equation}
The Greek index stands for 3d component whereas
the Latin for 2d.
Some of the necessary ingredients for the description of the sphere
is summarized in the Appendix for reference and notation convention.
We will consider the physical quantity  on one patch
first (the north patch for definiteness) and consider
the other patch when  necessary.
Hereafter, we drop the subscript $N$.

The Chern-Simons action for the gauge field $a_\mu$ is
given as
\begin{equation}
S_g = {\mu \over 2} \int d^3x
\sqrt g \epsilon^{\mu \nu \rho}
a_\mu \partial_\nu a_\rho
\label{II.1}
\end{equation}
where $d^3x   \sqrt g$ is the invariant volume element
and $\epsilon^{\mu\nu\rho}$ is the anti-symmetric Levi-Civita
tensor.
Let us introduce
the (pseudo-)invariant
$e^{\mu \nu \rho}= e_{\mu \nu \rho}$
with $e_{012} = 1$, which is related with $\epsilon^{\mu\nu\rho}$ as
\begin{equation}
\epsilon^{\mu\nu\rho}=
{1 \over \sqrt g} e^{\mu \nu \rho}\,, \qquad
\epsilon_{\mu\nu\rho}=
\sqrt g e_{\mu \nu \rho}
\end{equation}
using $g= \det (g_{\mu\nu})>0$.
Then the action is written as
\begin{equation}
S_g = {\mu \over 2} \int d^3x e^{\mu \nu \rho}
a_\mu \partial_\nu a_\rho
= {\mu \over 2} \int a\wedge da,
\end{equation}
which is independent of the metric tensor
as it should be.
The phase space variables of the gauge field are $a_1(x)$
and $a_2(x)$. $a_0(x)$ is considered as a Lagrange
multiplier. To quantize this system, we require the equal-time
commutation relations as
\begin{eqnarray}
&&[ a_1(x), a_2(x')] =
i {\hbar \over \mu} \delta^{(2)}(\vec x - \vec x')
\nonumber\\
&&[ a_i(x), a_0(x')] = 0
\label{a-comm}
\end{eqnarray}
and impose
the Lorentz gauge fixing condition,
\begin{equation}
\nabla_\mu a^\mu \equiv {1 \over \sqrt g}
\partial_\mu \sqrt g g^{\mu\nu} a_\nu = 0
\label{lorentz}
\end{equation}

To expand the gauge field in terms of modes,
we use the Hodge decomposition (see Eq. (\ref{A.9})),
rather than using the vector harmonics directly
\cite{edmonds}.
The gauge field is written as
\begin{equation}
a_i(x)  =  \partial_i \chi +
\epsilon_{ij} g^{jl} \partial_l \xi
\,.
\label{II.3}
\end{equation}
$\chi$ and $\xi$
are mode expanded using spherical harmonics $Y_{lm}(\Omega)$,
\begin{eqnarray}
\chi(x)&=& -{i \over \sqrt2} \sum_{l,m}{}'
[e^{-i\omega_l t} Y_{lm}(\Omega) \chi_{lm}
- e^{i\omega_l t} Y_{lm}^*(\Omega) \chi_{lm}^+]
\nonumber\\
\xi(x)&=&  \xi_B(x) +
\int d^2x' \sqrt{g(x')} G(\vec x, \vec x')
\zeta(x)
\label{mode}
\end{eqnarray}
where
\[
\zeta(x)= {1\over \sqrt2} \sum_{l,m}{}'
[e^{-i\omega_l t} Y_{lm}(\Omega) \zeta_{lm}
+ e^{i\omega_l t} Y_{lm}^*(\Omega) \zeta_{lm}^+].
\]
Here $\omega_l = l(l+1)$ and the prime on the summation stands
for the elimination of $l=0$ mode.
$G(\vec x, \vec x')$ is the rotation invariant Green's function
given in Eq. (\ref{A.6}). Each mode of the gauge field satisfies
the commutation relations,
\begin{equation}
[ \chi_{lm}, \zeta_{l'm'}^+]
= [ \zeta_{lm}, \chi^+_{l'm'}]=
{\hbar \over \mu} \delta_{ll'} \delta_{mm'}
\nonumber
\end{equation}
and commutes with the rest so that Eq. (\ref{a-comm})
is satisfied.
We added posssible background mode $\xi_B(x)$
in $\xi(x)$, which is
commuting with all the other modes.
According to the decomposition Eq. (\ref{II.3}),
we have
\begin{eqnarray}
\nabla_i a^i &\equiv&
{1 \over \sqrt g} \partial_i \sqrt g g^{ij} a_j
= - \nabla ^2 \chi
\nonumber\\
{1\over \sqrt g} b &\equiv&
- \epsilon^{ij}\partial_i a_j
= - \nabla ^2 \xi.
\end{eqnarray}
Therefore, the background mode $\xi_B$ can
accommodate a constant magnetic flux.
$a_0(x)$ can be written in consistency with the gauge fixing
condition Eq. (\ref{lorentz}) as
\begin{equation}
a_0(x)= \lambda +
 {1\over \sqrt2} \sum_{l,m}{}'
[e^{-i\omega_l t} Y_{lm}(\Omega) \xi_{lm}
+ e^{i\omega_l t} Y_{lm}^*(\Omega) \xi_{lm}^+].
\label{II.5}
\end{equation}
$\lambda$ is a space-time independent mode,
which is still to be gauge fixed.
(We also remark that the $\chi$ mode in Eq (\ref{II.3})
has residual gauge degree of freedom \cite{cr})

The relation between the gauge fields on either patch
is easily seen if we note the property of $\chi$ and $\zeta$
in Eq. (\ref{mode}).
Let us denote $\chi_{(N)}(P_N)$ and $\zeta_{(N)}(P_N)$
as the north-patch description of $\chi$ and $\zeta$ at
$\Omega$
respectively on the sphere.
On the south-patch, they are denoted as
$\chi_{(S)}(P_S)$ and $\zeta_{(S)}(P_S)$.
For the overlapping region,
the projected point $P_S(=z_S)$ of $\Omega=(\theta, \phi)$
to the south-patch
is to be identified as $\tilde P_N(=-1/ z_N)$
on the north-patch (FIG.\ref{fig.2});
$P_S \equiv  \tilde P_N$ on the north patch for the
overlapping region.
Then $\tilde P_N$ becomes the projected point of
$\tilde \Omega = (\pi - \theta, \pi - \phi)$.
Noting that
$
Y_{lm}(\tilde \Omega ) =
(-1)^l Y_{lm}^* (\Omega)
= (-1)^{l+m} Y_{l,-m}(\Omega)
$
we re-express $\chi_{(S)}(\tilde P_N)$
and $\zeta_{(S)}(\tilde P_N)$
as
\begin{eqnarray}
\chi_{(S)}(\tilde P_N)
&=&- {i \over \sqrt2} \sum_{l,m}{}'
\left[
e^{-i\omega_l t} Y_{lm}(\Omega )(-1)^{l+m} \chi_{(S)l,-m}
-e^{i\omega_l t} Y_{lm}^* (\Omega )(-1)^{l+m} \chi_{(S)l,-m}^+
\right]
\nonumber\\
\zeta_{(S)}(\tilde P_N)
&=& {1\over \sqrt2} \sum_{l,m}{}'
\left[
e^{-i\omega_l t} Y_{lm}(\Omega )(-1)^{l+m} \zeta_{(S)l,-m}
+e^{i\omega_l t} Y_{lm}(\Omega )(-1)^{l+m} \zeta_{(S)l,-m}^+
\right]
\label{II.6}
\end{eqnarray}
With the identification,
\begin{eqnarray}
\chi_{(N)lm}&=&(-1)^{l+m} \chi_{(S)l,-m},\quad
\chi_{(N)lm}^+=(-1)^{l+m} \chi^+_{(S)l,-m}
\nonumber\\
\zeta_{(N)lm}&=&(-1)^{l+m} \zeta_{(S)l,-m},\quad
\zeta_{(N)lm}^+=(-1)^{l+m} \zeta^+_{(S)l,-m},
\label{modeid}
\end{eqnarray}
we have
\begin{equation}
\chi_{(S)}(\tilde P_N)= \chi_{(N)}(P_N),
\quad \zeta_{(S)}(\tilde P_N) = \zeta_{(N)}(P_N).
\label{II.7}
\end{equation}
The sign reversal of $m$ arises because the rotation around
$\hat k$ on one patch corresponds to the counter-rotation
in the other. (See FIG.\ref{fig.1}).
Representing the gauge field in complex notation  as
\begin{equation}
a = a_1 - ia_2 = 2 \partial (\chi - i \xi),
\quad
\bar a = a_1 + ia_2 = 2\bar \partial(\chi + i \xi)
\label{II.9}
\end{equation}
we have the relation in the absence of the background
mode ($\xi_B=0$),
\begin{equation}
a_{(N)} (P_N) =  z_{S}^2\,\, a_{(S)}(\tilde P_N)
\label{II.10}
\end{equation}
using $\partial_{(N)} =  z_S^2 \,\, \partial_{(S)}$.
The presence of background mode will, however,  change the relation
in Eq.(\ref{II.10}), which will be important in the presence of
matter field.

\section{Gauged Non-relativistic matter field }

Let us couple the Chern-Simons gauge field to a non-relativistic
fermionic matter field $\psi(x)$. Its action is given
(on the north-patch) as
\begin{equation}
S_m = \int d^3x \sqrt g  \psi^+ i\hbar D_0 \psi
-{1 \over 2m}\int d^3x (i\hbar D_i \psi)^*(i\hbar D_i \psi)
\label{III.1}
\end{equation}
in addition to the gauge part $S_g$ in Eq. (\ref{II.1}).
Here $ i\hbar D_\mu \psi
\equiv (i\hbar \partial_\mu - {e\over c} a_\mu)\psi$.
The matter field satisfies the equal-time anti-commutation
relation,
\begin{eqnarray}
\{\psi(x), \psi^+(x')\}
&=& {1\over \sqrt g} \delta^{(2)}(\vec x -\vec x')
\nonumber\\
\{\psi(x), \psi(x')\}&=&0\,.
\label{III.2}
\end{eqnarray}
The hamiltonian is given as
\begin{equation}
H = \int d^2x \{ \mu a_0 \Gamma +
{1\over 2m} (i\hbar D_i \psi)^* (i\hbar D_i \psi)\},
\label{III.3}
\end{equation}
whose form is similar to that on a plane.
$\Gamma$ is the gauge generator given as
\begin{equation}
\Gamma = b + {e\over \mu c} \sqrt g J_0,
\label{constr}
\end{equation}
where $\sqrt g J_0$ is the number density
operator, $\, J_0 = \psi^+ \psi$.

Since the gauge field has no dynamics, we may decouple the gauge
field from the matter field by solving the constraint
and get the effective hamiltonian.
Instead of solving the constraint directly, we
re-express the hamiltonian
and the gauge generator in term of canonically
transformed new fields as done in Ref. \cite{swanson,cr}
\begin{eqnarray}
\Gamma &=& b^{(1)}(x) + {e \over \mu c} {Q \over 4\pi}
\sqrt {g(x)}
\nonumber\\
H &= & \int d^2x \{ \mu a_0 \Gamma +
{1\over 2m} (i\hbar D_i^{(1)} \psi^{(1)})^*
(i\hbar D_i^{(1)} \psi^{(1)})\}.
\label{III.4}
\end{eqnarray}
where $Q$ is the number operator,
$Q = \int d^2x \sqrt{g(x)}J_0(x)$.
Here the old fields are related with the new ones as
\begin{eqnarray}
a_i(x)  &=& V_1(t) a_i^{(1)}(x) V_1(t)^+
= a_i^{(1)}(x) - {e \over \mu c} e_{ij} \partial_j
\int d^2x' \sqrt{g(x')} G(\vec x, \vec x') J_0(x')
\nonumber\\
\psi(x) &=& V_1(t) \psi^{(1)}(x)V_1^+(t)
= \psi^{(1)}(x) \exp \{-i{e\over \hbar c}
\int d^2x'G(\vec x, \vec x')
\partial_i'
a_i^{(1)}(x')\}
\label{III.5}
\end{eqnarray}
where
\begin{equation}
V_1(t) = \exp \{ i{e\over \hbar c}
\int \int d^2x d^2x' \sqrt{g(x)}
J_0(x) G(\vec x, \vec x') \partial_i'
a_i^{(1)}(x')\}.
\end{equation}
The hamiltonian $H$ is normal-ordered such that
$\psi^{(1)+}$ lies in front of
the effective covariant derivative $D_i^{(1)}$. More explicitly,
\begin{equation}
(i\hbar D_i^{(1)} \psi^{(1)})^*
\equiv
\psi^{(1)^+} (i \hbar D_i)^+
\equiv
-i\hbar \partial_i\psi^{(1)+}
-{e\over c} \psi^{(1)+} a_i^{[1]eff}
\label{III.6}
\end{equation}
where
$$
a_i^{[1]eff}(x)= e_{ij} \partial_j
\{{e\over 4\pi\mu c} \ln (1 + \vec x^2)
- { e \over \mu c} \int d^2x \sqrt {g(x')} G(\vec x, \vec x')
J_0(x')
+ {1 \over 4\pi} \int d^2x' \ln (\vec x - \vec x')^2 b^{(1)}(x')\}
$$
It should be noted that $b^{(1)}(x)$ rather than $a_i^{(1)}(x)$ appears
in $a^{[1]eff}(x)$, which is the chief merit of this transformation
and the regularization
$
e_{ij}\partial_j \ln (\vec x - \vec x')^2|_{\vec x = \vec x'} = 0
$ is assumed.

Here, the gauge generator $\Gamma$ contains
the number operator
$Q$ as well as $b^{(1)}$.
We have not transformed away
$Q$ in the gauge generator
$\Gamma$ as in the plane case
due to possible conflict with $SO(3)$ rotation
(M\"obius transformation in complex notation)
invariance of the system.
To investigate the role of the
rotation symmetry more clearly,
let us introduce another
canonical fields which  get rid of $Q$ from $\Gamma$,
\begin{eqnarray}
a_i^{(1)}(x) &=& V_2 a_i^{(2)}(x) V_2^+
= a_i^{(2)}(x) - {eQ \over 4\pi \mu c}
e_{ij} \partial_j\ln (1 + \vec x^2)
\nonumber\\
\psi^{(1)}(x) &=& V_2 \psi^{(2)}(x)V_2^+
= \psi^{(2)}(x) \exp -i\eta_0 .
\label{III.7}
\end{eqnarray}
$V_2 = \exp \{-i Q \eta_0\}$
and $\eta_0$ is a conjugate
to $\xi_0$ of the background mode $\xi_B(x)$
defined as
$\xi_B(x) \equiv \xi_0 {e\over 4\pi\mu c} \ln(1+\vec x^2)$,
\begin{equation}
[\eta_0, \xi_0] =i
\label{III.8}
\end{equation}
Then the hamiltonian and the gauge generator in Eq.(\ref{III.4})
is written as
\begin{eqnarray}
\Gamma &=& b^{(2)}
\nonumber\\
H &=& \int d^2x \{\mu a_0 b^{(2)} + {1\over 2m}
(i\hbar D_i^{(2)}\psi^{(2)})^*
(i\hbar D_i^{(2)}\psi^{(2)})
\}
\label{III.9}
\end{eqnarray}
where
\begin{equation}
a_i^{[2]eff}(x)=
e_{ij} \partial_j
\int d^2x' {\ln (\vec x - \vec x')^2 \over 4\pi}
\{- {e\over \mu c} \sqrt{g(x')}J_0(x') + b^{(2)}(x')\}.
\label{III.10}
\end{equation}
The gauge generator $b^{(2)}$ commutes with all the
other fields in this representation and annihilates the
physical Hilbert space. The gauge
field $a_i^{(2)}(x)$ is therefore, decoupled from the
matter sector and we may drop $b^{(2)}$ from
the hamiltonian.
The effective gauge field
$a_i^{[2]eff}$ turns out to be the
solution of the constraint $\Gamma = 0$
in Eq. (\ref{constr}).

For the $SO(3)$ rotation invariant system,
no point is regarded as special.
Therefore, we have to introduce another equivalent hamiltonian
on the south-patch.
Adopt the same form of the hamiltonian on the
south-patch as
in Eq. (\ref{III.9}),
and we have a relation between the gauge
field on each patch,
which is a modified version of Eq. (\ref{II.10}),
\begin{equation}
a_{(N)}^{[2]eff}(P_N) =  z_S^2
[a_{(S)}^{[2]eff}(\tilde P_N)
+ {eQ\over 2\pi \mu c}
i\partial_{(S)} \ln ({z_S \over \bar z_S})]
\label{III.11}
\end{equation}
with the assumption that
\begin{equation}
J_{0(N)} (P_N) = J_{0(S)}(\tilde P_N),
\label{III.12}
\end{equation}
which should be checked a posteriori.
Considering the relation between the area element
on each patch,
$d^2 x_N  =d^2 x_S /|z_S|^4 $,
we have the relation between
matter fields
up to a trivial constant phase factor,
\begin{equation}
\psi_{(N)} (P_N) = ({ z_S \over \bar z_S})^{{\nu   \over 2}Q}
\,\, \psi_{(S)} (\tilde P_N)
\label{III.13}
\end{equation}
where $\nu = {e^2 \over 2\pi\mu c^2 \hbar}$.
The number density satisfies the condition in Eq. (\ref{III.12}).
However, the presence of the number operator
in the phase factor spoils this formalism
since the identification in Eq. (\ref{III.13}) will not
respect the anti-commutation relations given in Eq. (\ref{III.2}).

To avoid this obstruction, we present two ways out.
One is to simply abandon $SO(3)$
but allows only $SO(2)$ rotation invariance
around the axis through a point (e.g., the south-pole)
which is to be excluded from the manifold.
A sphere with the elimination of the one point is
topologically equivalent
to an infinite plane: We need only one coordinate-patch,
which removes the
subtlety. This system is best
described with the hamiltonian in Eq. (\ref{III.9}).

Another way is to maintain the $SO(3)$ invariance
but changes the way of applying the gauge constraint
on the physical state.
The problem with the (anti-)commutation relation
traces back to the presence of the number operator $Q$ attached to
$SO(3)$ rotation non-invariant term $\ln (1+\vec x^2)$ in
$a_i^{[2]eff}(x)$.
To cure this, we go back to the
hamiltonian and gauge generator
given in Eq.(\ref{III.4}).
Let us express $\Gamma$ as
\begin{equation}
\Gamma(x) =
(b^{(1)} + {e \over \mu c}{\sqrt{g(x)}\over 4\pi}N)
 + {e \over \mu c}{\sqrt{g(x)}\over 4\pi}(Q-N)
\label{III.14}
\end{equation}
where $N$ is a $c$-number integer,
which will be interpreted as the number of particles.
Let us introduce
similar canonical fields in Eq. (\ref{III.7}).
\begin{eqnarray}
a_i^{(1)}(x) &=& V_3 a_i^{(3)}(x) V_3^+
= a_i^{(3)}(x) - {eN \over 4\pi \mu c}
e_{ij} \partial_j\ln (1 + \vec x^2)
\nonumber\\
\psi^{(1)}(x) &=& V_3 \psi^{(3)}(x)V_3^+
= \psi^{(3)}(x)
\label{III.15}
\end{eqnarray}
where $V_3 = \exp \{-iN\eta_0\}$. In terms of this new fields,
we have the constraint
\begin{equation}
\Gamma = b^{(3)}
 + {e \over \mu c}{\sqrt{g(x)}\over 4\pi}(Q-N)
\nonumber
\end{equation}
and the effective gauge term
\begin{eqnarray}
a_i^{[3]eff}(x)= e_{ij} \partial_j
\{&-&
{e(N-1)  \over 4\pi\mu c} \ln (1 + \vec x^2)
- { e \over \mu c} \int d^2x' \sqrt {g(x')} G(\vec x, \vec x')
J_0(x')
\nonumber\\
&&
+ {1 \over 4\pi} \int d^2x' \ln (\vec x - \vec x')^2 b^{(3)}(x')\}
\end{eqnarray}

We constrain the physical space as
\begin{equation}
b^{(3)} |phys> = 0
\qquad
Q|phys> = N |phys>\,.
\label{III.17}
\end{equation}
Fixing the remnant gauge as
$\lambda = 0$ and dropping $b^{(3)}$, we have hamiltonian
as
\begin{equation}
H =  {1\over 2m}
\left|
(
i\hbar \partial_i+
e_{ij} \partial_j
\{{e^2(N-1)  \over 4\pi\mu c^2} \ln (1 + \vec x^2)
+ { e^2 \over \mu c^2} \int d^2x' \sqrt {g(x')} G(\vec x, \vec x')
J_0(x') \}
)
\psi^{(3)}
\right|^2
\label{III.18}
\end{equation}
The difference of this hamiltonian from the one given in
Eq. (\ref{III.9}) is clear: $Q$ is replaced by a $c$-number
$N-1$
(due to the normal-ordering of the operators,
$\psi^+$, $Q$ and $\psi$).
The equivalence between the matter fields on
the north and the south patch
is established by noting
\begin{equation}
\psi_{(N)} (P_N)
= ({ z_S \over \bar z_S})^{{\nu \over 2}(N-1)}\,\,
\psi_{(S)} (\tilde P_N)
\label{III.19}
\end{equation}
which respects the anti-commutation relation on either patch.
The transition function in Eq. (\ref{III.19}),
however, has the cut from the south to north pole. To
remove this cut, we need the Dirac quantization condition
similarly appearing in the magnetic monopole analysis,
\begin{equation}
(N-1) \nu  = n
\label{III.20}
\end{equation}
where $n$ is an arbitrary integer.
The factor $(N-1)$ rather than $N$ arises because
the particle does not
see its own flux.
This condition is noted in the braid group analysis
on the sphere \cite{thou}.
By making the gauge sector look trivial
(Eq. (\ref{III.17}))
except the possible quantum holonomy,
we have instead the  explicit interaction term due to the
magnetic monopole-like source inside the sphere.
(See Eq. (\ref{III.18}).
If we go back to the original picture,
$b^{(1)}$ in Eq. (\ref{III.14}) is seen to possess
the monopole source from the beginning
proportional
to the number of particles $N$.

One may also consider the case when $N$ is non-positive.
For example, when  $N=0$
the Hilbert space contains no particles at all in the above
analysis.
To make it physically  interesting, one may
introduce multi-species of the matter fields  where each species
has different coupling.
We present
the simplest choice, i.e.,
two species of matter fields
with equal but opposite statistical coupling. Then the
Hilbert space with $N=0$ consists of states with arbitrary number
of particles and equal number of ``anti-particles''.
The hamiltonian can be written
as
\begin{equation}
H =  {1\over 2m_{(+)}}\int d^2x
|i\hbar D_{i(+)}^{(1)}\psi_{(+)}^{(1)}(x)|^2
+  {1\over 2m_{(-)}}\int d^2 x
|i\hbar D_{i(-)}^{(1)}\psi_{(-)}^{(1)}(x)|^2
\label{III.21}
\end{equation}
The derivative operator is given as
\begin{equation}
D_{i(\pm)} ^{[1]}
= i \hbar \partial_i \mp
{e  \over \mu^2 c}
e_{ij} \partial_j
\{{1  \over 4\pi} \ln (1 + \vec x^2)
- \int d^2x \sqrt {g(x')} G(\vec x, \vec x')
J_0(x')\}
\nonumber
\end{equation}
where the charge density operator is defined as
$
J_0(x) = \psi^+ _{(+)} \psi_{(+)}
- \psi^+ _{(-)} \psi_{(-)}
$.
The realtion of the matter field on each patch is given as
\begin{equation}
\psi_{(N)(\pm)} (P_N)
= ({ z_S \over \bar  z_S})^{\mp {\nu  \over 2}}\,\,
\psi_{(S)(\pm)} (P_S)
\label{III.22}
\end{equation}
which recalls the Dirac quantization condition in
Eq. (\ref{III.20}) with $N=0$;
\begin{equation}
\nu  =n
\label{III.23}
\end{equation}
Quite straight-forwardly, one can construct
the system with
negative $N$ introducing multi-species.

\section{Anyon equation}

Suppose there are $N$ identical particles on a sphere.
The dynamics of the particles can be described
in two ways as given in Sec. III,
which is distinguished by the rotation invariance.
With only $SO(2)$ invariance, the particles obey the
dynamics with the hamiltonian in Eq. (\ref{III.9}).
The statistical flux need not be quantized
similarly as in the plane.
On the other hand, $SO(3)$ invariance leads to the
flux quantization Eq. (\ref{III.20}),
and the dynamics is described
by the hamiltonian Eq. (\ref{III.18}).
The Schr\"odinger equation
can be obtained from the Heisenberg equation
for the corresponding matter field and its application to the
$N$-body wave function.
For the case with $SO(2)$ invariance,
we obtain the Heisenberg equation,
\begin{eqnarray}
i\hbar {\partial \psi(x) \over  \partial t} &=&
{1 \over \sqrt g} (i \hbar D_i^{(2)})^2 \psi (x)
\nonumber\\
&+& ({e\over c})^2 \int d^2 x' \psi^+(x') \psi(x')
(R_i(\vec x', \vec x))^2 \psi(x)
\nonumber\\
&-&({e\over c})\int d^2x' \psi^+(x')
[(i\hbar D_i{'}^{(2)})^+  + i\hbar D_i {'}^{(2)}]
\psi(x') R_i(\vec x', \vec x) \psi(x)
\label{IV.1}
\end{eqnarray}
where
$R_i(\vec x', \vec x) \equiv - {e\over 4 \pi \mu c}
e_{ij}\partial'_j\ln(\vec x- \vec x')^2$.
Defining the $N$-particle wavefunction as
\begin{equation}
\Phi(N) = <0|\prod_{p=1}^N \psi( x_p)|N>
\label{IV.2}
\end{equation}
where $|N>$ is $N$-particle Heisenberg state,
we have
\begin{equation}
i\hbar {\partial \Phi(N) \over \partial t} =
\sum_{p=1}^N {1\over 2m} {1\over \sqrt {g(x_p)}}
(i\hbar \partial_i^{(p)} + {\cal A}_i(x_p))^2
\Phi(N)
\label{IV.3}
\end{equation}
where
\begin{equation}
{\cal A}_i(x_p)= {e^2 \over 4 \pi \mu c^2}
e_{ij}\partial_j^{(p)} \ln \{\prod_{q (\ne p)}
(\vec x_p -\vec x_q)^2 \}.
\label{IV.4}
\end{equation}
${\cal A}_i(x_p)$ satisfies the constraint,
\begin{equation}
e_{ij}\partial_j^{(p)}
{\cal A}_j(x_p) =
- {e^2 \over 4\pi \mu c^2} \sum_{q (\ne p)}
\delta^{(2)}(\vec x_p - \vec x_q)\,.
\label{IV.6}
\end{equation}

When there is a single particle ($N=1$), the particle moves free
(${\cal A}_i = 0$).
For $ N \ge 2$,
the wavefunction should satisfy
the hard-core boundary condition
between particles and also
vanish at the south-pole.
Since  there is no statistical flux quantization,
there exist cuts
from the south-pole to the particle positions.
In addition, one can transform the equation
to a free form through a singular transformation,
\begin{equation}
\tilde \Phi (N) = \prod_{p > q}
({z_p -z_q \over \bar z_p - \bar z_q})
^{\nu \over 2 }
\Phi (N)
\label{IV.5}
\end{equation}
where $z_p =x_p^{(1)} + i x_p^{(2)}$.
$\tilde \Phi(N)$ is multi-valued
under the exchange of particles
whereas
$\Phi(N)$ is anti-symmetric.

For the $SO(3)$ invariant case,
using the wavefunction in Eq. (\ref{IV.2})
and Hamiltonian Eq. (\ref{III.18}),
we have the Schr\"odinger equation of the form in Eq. (\ref{IV.3})
where ${\cal A}_i(x_p)$ is now given as
\begin{equation}
{\cal A}_i(x_p) =
\left\{
\begin{array}{ll}
0 & \mbox{for $N=1$}\\
{e^2\over \mu c^2} e_{ij} \partial_j^{(p)}
\{{N-1 \over 4\pi} \ln (1 + \vec x^2)
+ \ln \prod_{q(\ne p)}G(\vec x_p, \vec x_q)\}
&\mbox{for $N \ge 2$ }
\end{array}
\right.
\end{equation}
which remarkably turns out to be the same
as the one given in Eq. (\ref{IV.4}).
The distinctive point lies in that
the statistical
flux is quantized and
one has only to apply the hard-core boundary condition
between particles on the wavefunction.
(There is no cut from the particle position to the south-pole.)
The same Schr\"odinger equation is also
obtained by Comtet {\it et. al.} in \cite{lee}.

The wavefunction
$\tilde \Phi(N)$ is easy to obtain
when $\nu$ is an even integer,
which is ordinary anti-symmetric
product of the one-particle wavefunctions.
However, for odd $\nu$,
$\tilde \Phi$ is to be symmetric
satisfying the hard-core condition,
which is not obtained merely by symmetrization
of the products of the one-particle
wavefunctions.
For general value of $\nu$, the wavefunction is not known yet.

Schr\"odinger equation for two species can be
obtained similarly.
Using the hamiltonian in Eq. (\ref{III.21}),
one obtains the Heisenberg equation for the field,
\begin{eqnarray}
&&i\hbar {\partial \psi_{(\pm)}(x) \over  \partial t} =
\pm {1 \over 2m_{(\pm)}\sqrt g}
(i \hbar D_{i(\pm)}^{(1)})^2 \psi_{(\pm)} (x)
\nonumber\\
&&\,\, + ({e\over c})^2 \int d^2 x'
\left[
{\psi^+_{(+)}(x') \psi_{(+)}(x') \over 2 m_{(+)}}
+{\psi^+_{(-)}(x') \psi_{(-)}(x') \over 2 m_{(-)}}
\right]
(K_i(\vec x', \vec x))^2 \psi_{(\pm)}(x)
\nonumber\\
&&\,\, \mp ({e\over 2m_{(+)} c})\int d^2x'
\left[
\psi^+_{(+)}(x') (i\hbar D_{i(+)}^{\prime (1)})^+
\psi_{(+)}(x')
+\psi^+_{(+)}(x') (i\hbar D_{i(+)}^{\prime (1)} \psi_{(+)}(x') )
\right]
K_i(\vec x', \vec x) \psi_{(\pm)}(x)
\nonumber\\
&&\,\, \pm ({e\over 2 m_{(-)}c})\int  d^2x'
\left[
\psi^+_{(-)}(x') (i\hbar D_{i(-)}^{\prime (1)})^+
\psi_{(-)}(x')
+\psi^+_{(-)}(x')
(i\hbar D_{i(-)}^{\prime (1)} \psi_{(-)}(x'))
\right]
K_i(\vec x', \vec x) \psi_{(\pm)}(x)
\label{IV.7}
\end{eqnarray}
where
$i\hbar D_i^{(1)}
\equiv -i\hbar \partial_i -{e\over c} a_i^{[1]eff}$
and
$K_i(\vec x', \vec x) \equiv - {e\over 4 \pi \mu c}
e_{ij}\partial'_j G(\vec x, \vec x')$.
Defining the $N$-particle wavefunction as
\begin{equation}
\Phi(N;N) \equiv
<0| \prod_{p=1}^N \psi_{(+)}(\vec x_p)
\psi_{(-)}(\vec y_p)|N;N>,
\label{IV.8}
\end{equation}
we have the Schr\"odinger equation for
$\Phi(N;N)$ as follows.
\begin{eqnarray}
i\hbar {\partial \Phi(N;N) \over \partial t} =
\sum_{p=1}^N &\{&
{ 1\over 2m_{(+)}} {1\over \sqrt {g(x_p)}}
(i\hbar \partial_i^{x_p} + {\cal A}_i(x_p))^2
\nonumber\\
&+& {1\over 2m_{(-)}} {1\over \sqrt {g(y_p)}}
(i\hbar \partial_i^{y_p} + {\cal B}_i(y_p))^2
\}
\Phi(N;N)
\label{IV.9}
\end{eqnarray}
where
\begin{eqnarray}
{\cal A}_i(x_p) &=&{e^2 \over 4 \pi \mu c^2}
e_{ij}\partial_j^{x_p}
\ln{1 \over (\vec x_p - \vec y_p)^2}
\{\prod_{q (\ne p)}
{(\vec x_p -\vec x_q)^2
\over (\vec x_p -\vec y_q)^2}
\}
\nonumber\\
{\cal B}_i(y_p) &=&{e^2 \over 4 \pi \mu c^2}
e_{ij}\partial_j^{y_p}
\ln{1 \over (\vec x_p - \vec y_p)^2}
\{\prod_{q (\ne p)}
{(\vec y_p -\vec y_q)^2
\over (\vec y_p -\vec x_q)^2}
\}
\nonumber
\end{eqnarray}
One can put the Schr\"odinger equation in a free form if one makes
a singular gauge transform,
\begin{equation}
\tilde \Phi(N;N)=
\left[
(\prod_{p,q}{\bar z_p - \bar w_q
\over z_p -w_q})
(\prod_{p>q}
{(z_p -z_q)(w_p-w_q)\over
(\bar z_p - \bar z_q)(\bar w_p- \bar w_q)})
\right]^{\nu \over 2}
\Phi(N;N)
\label{IV.10}
\end{equation}
where $z_p = x_p^{(1)}+ ix_p^{(2)}$
and
$w_p = y_p^{(1)}+ iy_p^{(2)}$.
$\tilde \Phi(N;N)$ may change its statistics
when two particles of the same species interchange
each other by the factor
$(-1)^\nu$ which is $\pm1$ according to the Dirac
quantization condition in Eq. (\ref{III.23}).

\section{External magnetic source}

Suppose one introduces external magnetic source,
whose field is constant on the surface and is
perpendicular to the surface, magnetic monopole.
Then the hamiltonian in Eq. (\ref{III.1}) is modified
such that the gauge field is replaced by $a_i +
A_i^{ext}$
where
$A_i^{ext} = {\phi \over 4 \pi} e_{ij} \partial_i
\ln (1 + \vec x^2)$
(or $b^{ext} = {\sqrt g \over 4\pi} \phi$).
For the $SO(3)$ invariance case, one can re-express the hamiltonian
as in Eq. (\ref{III.18}),
where the covariant derivative is shifted by
$- {e\over c} A_i^{ext}$.
The compatibility of the two patch description
leads to the Dirac quantization condition.
\begin{equation}
(N-1) \nu - k = m
\label{DiracV}
\end{equation}
where $k = {e\phi \over hc}$.

Also the Schr\"odinger equation for $N$ particles
is given as in Eq. (\ref{IV.3})
where ${\cal A}_i (x_p)$ is replaced by
${\cal A}_i (x_p) - {e\phi \over 4 \pi c} e_{ij} \partial_j
\ln (1+ \vec x_p^2)$,
as is  commonly found in the literature.
In case $k$ is strong enough such that $m\le 0$,
one can easily construct the ground states
and find their degeneracy.
Define
\begin{equation}
i \hbar {\cal D} = i \hbar ({\cal D}_1 - i {\cal D}_2),
\quad
i \hbar \bar {\cal D} = i \hbar ({\cal D}_1 + i {\cal D}_2),
\end{equation}
we have the Schr\"odinger equation for $N$ particles
\begin{equation}
i\hbar {\partial \Phi(N) \over \partial t} =
{ \hbar^2 \over 2m}
\sum_{p=1}^N
{1\over \sqrt {g(x_p)}}
\left( - {\cal D}\bar {\cal D} (x_p)
+ {k \over 2} \sqrt{g(x_p)} \right)
\Phi(N)\,.
\label{Vschr}
\end{equation}
The ground state is obtained
\cite{li} when
$
\bar {\cal D} (x_p) \Phi (N) = 0
$
whose solution has the form,
\begin{equation}
\Phi (N) = \left \{ \begin{array}{ll}
\prod_{p<q}(z_p - z_q)
\prod_{p<q}|z_p - z_q|^\nu
\prod_p (1+|z_p|^2)^{-{k \over 2}}\,\,
S(z);
& 0 \le {\nu \over 2}< 1 \\
\prod_{p<q}(\bar z_p - \bar z_q)^{(2r+1)}
\prod_{p<q}|z_p - z_q|^{(\nu-4r -2)}
\prod_p (1+|z_p|^2)^{-{k \over 2}}\,\,
S(z); \quad
& 2r+1  \le {\nu \over 2}< 2r +3
\end{array}
\right.
\label{ground-sol}
\end{equation}
where $r$ is a non-negative integer.
$S(z)$ is a symmetric holomorphic polynomial
with degree $P_s$.
The normalizability of $\Phi(N)$ restricts $S(z)$:
The power of $|z_p|$
should satisfy
$$
\left \{ \begin{array}{ll}
(N-1) (\nu+1) - k  + P_s \le 0;\quad
& 0 \le {\nu \over 2} < 1 \\
(N-1) (\nu-2r-1) - k  + P_s \le 0 ; \quad
& 2r+1  \le {\nu\over 2} < 2r+3
\end{array}
\right.
$$
When $S(z)=1 $ and $N \to \infty$,
filling factor $\alpha \equiv N / k$
is given as
\begin{equation}
\alpha = \left \{ \begin{array}{ll}
{1 \over \nu +1}; \quad & 0 \le {\nu\over 2} < 2
\\
{1 \over \nu-(2r+1)};
\quad & 2r+1 \le {\nu\over 2} < 2r+3\,.
\end{array}
\right.
\label{filling}
\end{equation}
(For the bosonic wavefunction,
$\Phi(N)$ in Eq. (\ref{ground-sol}) is to be symmetrized and
the corresponding filling factor is to be modified
from Eq. (\ref{filling}).)

The Schr\"odinger equation becomes the ordinary
monopole system after the singular gauge transformation
\begin{eqnarray}
i\hbar {\partial \tilde \Phi(N) \over \partial t} &=&
{ 1 \over 2m} \sum_{p=1}^N
{1\over \sqrt {g(x_p)}}
(i\hbar \partial_i^{(p)} - {e \over c} A_i^{ext}(x_p))^2
\tilde  \Phi(N)
\nonumber\\
&=& {\hbar ^2 \over 2m} \sum_{p=1}^N
\{ {1 \over 2}(L_{p+} L_{p-} + L_{p-} L_{p+})
+ (L_{p3})^2 - ({k \over 2})^2 \}
\tilde  \Phi(N)
\label{schr-mono}
\end{eqnarray}
where
$
L_{p+} =  - \bar \partial^{(p)}
- z_p^2 \partial^{(p)} + {k \over 2} z_p
$,
$
L_{p-} =  \partial^{(p)}
+\bar  z_p^2 \bar \partial^{(p)} + {k \over 2} \bar z_p
$,
$
L_{p3} =
 z_p \partial^{(p)}
- \bar z_p\bar \partial^{(p)}
- {k \over 2} z_p.
$
$L_{p}$'s are $su(2)$ generators
\cite{wuyang},
\begin{eqnarray}
[ L_{p+} , L_{q-} ] &=& 2 L_{p3} \delta_{pq}
\nonumber\\{}
[ L_{p3} , L_{q \pm} ] &=& \pm  L_{p \pm } \delta_{pq}.
\end{eqnarray}
Without the statistical interaction, one can obtain the
spectrum and degeneracy using the $su(2)$ representation.
On the other hand, since $\tilde \Phi$ is multi-valued
due to the statistical interaction, the usefulness
of $su(2)$ algebraic structure is not clear.
Although the ground state solution in Eq. (\ref{ground-sol})
is the eigenstate of $\sum_p L_{p3}$ and $H$,
it is not annihilated by $L_{p-}$
($\sum_p L_{p-}$ does annihilate the state but it does not
help much).

\section{Summary and Discussion}

We have investigated the role of the rotation symmetry  to  the  anyon
equation employing the canonically transformed new fields  in  the  CS
hamiltonian. To maintain the $SO(3)$ rotation invariance, one  has  to
restrict the Hilbert space to the one with definite charge.
The  consistency  for  the  two  patch  description  requires  the  CS
coefficient quantized,
which has the same consequencies as in the effect
due to the monopole source.

On  the  other  hand,  if  one  allows  $SO(2)$
rotation invariance only, then one may eliminate one  point  from  the
sphere and the manifold is topologically equivalent  to  the  infinite
plane.
The Hilbert space need not be restricted to the definite charge states
and the CS coefficient does not have to be quantized.
The Schr\"odinger equation is, however,
remarkably the same for  both  the cases
while the  boundary  conditions  to   be   applied   on   the
wavefunction differ.

The analysis on the sphere is equally applicable to the
case on a torus.
The appearance of  non-commuting zero-modes of the
gauge field due to the harmonic form
makes the analysis complicated in the intermediate
steps. However, once we get the effective gauge terms
for the matter field, the analysis is straight-forward.

If  one  identifies the fundamental domain of the torus
as the rectangular lattice (see FIG.\ref{fig.3}),
then the CS theory can be made
periodic both in the direction:
The translation operators commute each other.
In this case, there is no
Dirac quantization condition.
On the other hand, suppose
one has the freedom to choose a
parallelogram (see FIG.\ref{fig.4}) as the
unit cell (modular transformation) \cite{lechner},
then the translation operators do not commute each other.
The trouble lies in the effective gauge term,
which is given as
\begin{equation}
A_i^{eff}(x)=
-{e\over \mu c} \epsilon_{ij} \partial_j
\int d \vec y G_p(\vec x, \vec y) J_0(y)
-{e\over \mu c} \epsilon_{ij} \partial_j
\int d \vec y G_{np}(\vec x, \vec y) J_0(y)
\end{equation}
where
\begin{eqnarray}
G_p(\vec x, \vec y)&=&
{1\over 4\pi}\ln
\left|
{\theta_1(z|\tau) \over \theta'_1(0|\tau)}
\right|^2
-{(x_2 - y_2)^2 \over 2 L_1 L_2}
\nonumber\\
G_{np} (\vec x, \vec y)
&=& {(x_2 -y_2)^2 \over 2L_1L_2}\,.
\nonumber
\end{eqnarray}
(Here we follow the same notation given in \cite{cr}.)
$G_p$ is periodic under the translation along the modular
transformed unit cell,
whereas $G_{np}$ is not.
The translation results in the additional term depending on the
total charge operator $Q$,
\begin{equation}
A_i^{eff} (x) \to
\left\{
\begin{array}{ll}
A_i^{eff}(x)&
\mbox{when $x^1 \to x^1 + L_1$} \\
A_i^{eff}(x) - {e \over \mu c} {Q \over L_1}
\delta_{i1} &
\mbox{when $x^2 \to x^2 + L_2$ .}
\end{array}
\right.
\end{equation}
This result is similar to the one in sphere
Eq. (\ref{III.11}).

To fix this, one can have two choices.
One is to abandon the modular invariance:
One requires the square lattice as the fundamental domain
and gives the periodic condition at the edge.
The other way is to maintain the modular invariance
but restricts the Hilbert space to the one
with definite charges. Following the similar produre done
on the sphere,
one can obtain the effective gauge term,
\begin{equation}
A_i^{eff}(x)=
- (N-1) \tilde x_i
-{e\over \mu c} \epsilon_{ij} \partial_j
\int d \vec y G_p(\vec x, \vec y) J_0(y)
+ \int d \vec y \tilde y_i J_0(y)
\end{equation}
where
$\tilde x_1 = {e \over \mu c} {x^2 \over L_1 L_2}$
and $\tilde x_2 = 0$.
In this case, we have
\begin{equation}
A_i^{eff} (x) \to
\left\{
\right.
\begin{array}{ll}
A_i^{eff}(x) &
\mbox{when $x^1 \to x^1 + L_1$} \\
A_i^{eff}(x) - {e(N-1) \over \mu c L_1}\,
\delta_{i1}&
\mbox{when $x^2 \to x^2 + L_2$\,. }
\end{array}
\end{equation}
Then the  field operator has the translation property
\begin{eqnarray}
\psi(\vec x + L_2 \hat e_2) &=&
\psi (\vec x)
\exp\left\{
{ie^2 (N-1)x^1\over \hbar \mu c^2L_1}
\right\}
\nonumber\\
\psi(\vec x + L_1 \hat e_1) &=& \psi(\vec x)\,.
\label{bound}
\end{eqnarray}
The consistency condition requires the same  quantization
condition as in the sphere Eq. (\ref{III.20}).
The many-anyon Schr\"odinger equation on the torus
obtained in \cite{cr} is also applied to this case.
The boundary condition on the wavefunction is,
however, to be assigned in accordance with
the appropriate symmetry in accordance with
Eq. (\ref{bound}).

Finally, it would be challenging
to extend this analysis
to the non-abelian CS theory and demonstrate
the anyon equation obtained in \cite{verlinde}
from the second-quantized field theory.

\acknowledgments

The authors have  benefit from the discussion with Dr. K. Cho.
This work is supported in part by KOSEF No. 931-0200-030-2,
by Basic Science Research Institute program
(BSRI-94-2434) Ministry of Education, and
by SRC program through Seoul National University.

\appendix
\section{}
We summarize some of the useful formulae on a sphere.
The line element on the sphere is written as
\begin{eqnarray}
(ds)^2 &=& (d\theta)^2 + \sin ^2 \theta (d\phi)^2
\nonumber\\
&=& ({2 \over 1+ \vec x^2})^2
[(dx^1)^2 + (dx^2)^2]
\label{A.1}
\end{eqnarray}
where $\Omega=(\theta, \phi)$ and $P_N = (x^1, x^2)$ according to
FIG.\ref{fig.1}.
This holds for the north-patch
($0 \le \theta <\pi; |\vec x|<\infty$).
Similar formula can be expressed for the south-patch.
The invariant measure is given as
\begin{equation}
d \tau = \sin \theta d\theta d\phi
= \sqrt g dx^1 d x^2
\label{A.2}
\end{equation}
where $g$ is the determinant of the metric in space
2 dimension,
\begin{equation}
g_{ij} = - ({2 \over 1+ \vec x^2})^2
\delta_{ij}
= - \sqrt g \delta_{ij}
\end{equation}
($\delta_{11}=\delta_{22}=1$,
$\delta_{12} = \delta_{21}=0$).
We add the negative sign to $g_{ij}$ to make it consistent with
the 3-d metric whose signature is
($+, -,-$).
The 2-d laplacian is given as
\begin{equation}
\nabla^2 \equiv - {1\over \sqrt g}\partial_i
\sqrt g g^{ij} \partial_j=
{1\over \sqrt g} \delta_{ij} \partial_i\partial_j=
({1 + \vec x^2 \over 2})^2
(\partial_1^2 + \partial_2^2)\,.
\label{A.3}
\end{equation}
The 2-d delta function is written as
\begin{equation}
\delta^{(2)}(\Omega - \Omega')=
{1 \over \sqrt g}\delta
(x^1 - x^1{}')\delta (x^2 - x^2{}')
= {1\over \sqrt g}\delta ^{(2)}(\vec x-\vec x')
\label{A.4}
\end{equation}
The Green's function on the sphere is defined as
\begin{equation}
\nabla^2 G(\vec x, \vec x') = {1 \over \sqrt g}
\delta^{(2)}(\vec x - \vec x') - {1 \over 4 \pi}
\label{A.5}
\end{equation}
where the constant term in the RHS is due to the lack of
zero-mode contribution to the Green's function.
Explicitly, $G(\vec x, \vec x')$ can be written as
\begin{equation}
G(\vec x, \vec x') = {1 \over 4\pi}
\ln \left\{ {(\vec x- \vec x')^2 \over
(1+\vec x^2)(1+ \vec x'^2)}
\right\}
\label{A.6}
\end{equation}

We write down some of the properties of $G(\vec x, \vec x')$.
\begin{enumerate}
\item
$G(\vec x, \vec x')$ is invariant under the rotation of the sphere,
which corresponds to the M\"obius transformation
in the complex notation,
$z = x^{(1)} + ix^{(2)} \to
{a z + b \over - \bar b z + \bar a}$
where $a \bar a + b \bar b \ne 0$.
\item
$(\partial_1\partial_1' +
\partial_2\partial_2') G(\vec x, \vec x') =
-\delta ^{(2)} (\vec x - \vec x')$,
which is checked by explicit calculation.
\item
$\int d^2 x \sqrt {g(x)} G(\vec x, \vec x')=
\int d^2 x \sqrt {g(x)} G(\vec x, 0)= -1$
due to the rotation invariance.
\item
$G(\vec x, \vec x') = G(\vec x', \vec x) $.
\end{enumerate}

The gauge field on the sphere can be decomposed
into the gradient part and the
curl part as in the plane.
For this purpose, we introduce
form notation.
The gauge one form is given as
$a = a_i dx^i$. The measure in Eq. (\ref{A.2})
is expressed in two form  and
is given as $d\tau = \sqrt g dx^1 \wedge dx^2$.
In terms of anti-symmetric tensor $\epsilon_{ij}
(\equiv \epsilon_{0ij})$,
we can write this area form as
$d \tau = {1 \over 2} \epsilon_{ij} dx^i \otimes dx^j$,
where $\epsilon_{ij} = \sqrt {|g|} e_{ij}$
and $e_{ij}$ is a (pseudo) invariant  with
$e_{12}=e^{12}= -e^{21} =-e_{21}=1$.
The anti-symmetric tensor enables us to define
the duality (Hodge $*$) transformation
which transforms $r$-form into $2-r$ form,
\begin{equation}
*(dx^{\mu_1}\wedge \cdots \wedge dx^{\mu_r} )=
{1\over (2-r)!}
{\epsilon^{\mu_1 \cdots \mu_r}}_{\nu_{r+1} \cdots \nu_m}
(dx^{\nu_{r+1}}\wedge \cdots \wedge dx^{\nu_m})
\nonumber
\end{equation}
Explicitly, $*1 = \sqrt g dx^1 \wedge dx^2$,
$* dx^i = {\epsilon^i}_j dx^j
= g^{ij} \epsilon_{kj}dx^j
=- e_{ij}dx^j$,
$* \sqrt g dx^1 \wedge dx^2 = \sqrt g \epsilon ^{12}
=e_{12} =1$.

The gauge one form is decomposed (Hodge decomposition)
as
\begin{equation}
a = d \chi + d^+ (\xi d\tau)
\label{A.7}
\end{equation}
where $\chi$ and $\xi$ are zero-one form (function)
and $d^+$ is the adjoint operator defined as
\begin{equation}
d^+(\xi d\tau) = -* d *(\xi d \tau) =
-e_{ij}(\partial_j \xi) d x^i
\label{A.8}
\end{equation}
There is no harmonic form to the decomposition.
In component notation, the gauge field
is expressed as
\begin{equation}
a_i = \partial_i \chi  - e_{ij}\partial_j \xi
= \partial_i \chi + \epsilon_{ij} g^{jl}
\partial_l \xi \,.
\label{A.9}
\end{equation}

\begin{figure}
\caption{(a) North patch:
	 $\Omega$ on the sphere is projected to $P_N$
	 on the plane and the north pole to $O$.
	 The south pole is eliminated.
	 (b) South patch: $\Omega$ is projected to $P_S$
	 and the south pole to $O$. The north-pole is
	 eliminated.}
\label{fig.1}
\end{figure}

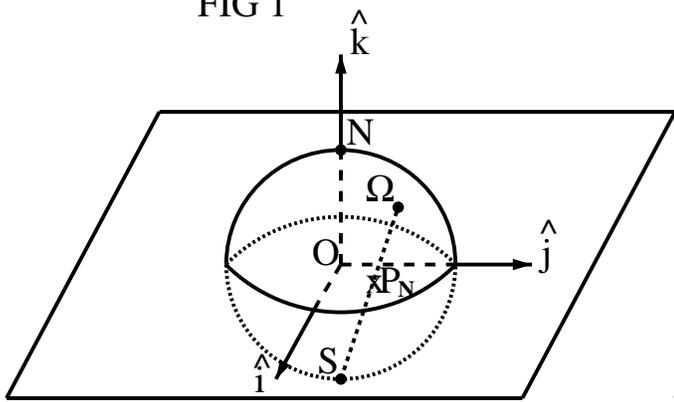
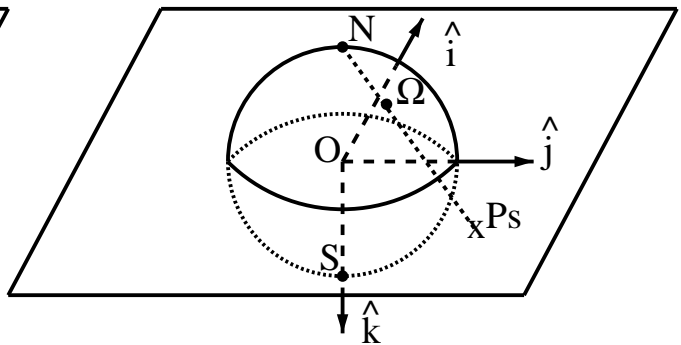
\begin{figure}
\caption{Identification of the north patch with the south.}
\label{fig.2}
\end{figure}

\begin{figure}
\caption{Rectangular unit cell
as the fundamental domain of the torus.}
\label{fig.3}
\end{figure}

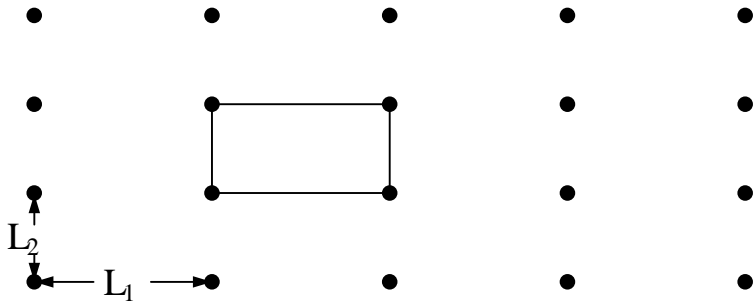
\begin{figure}
\caption{Two kinds of modular transformed unit cells.}
\label{fig.4}
\end{figure}

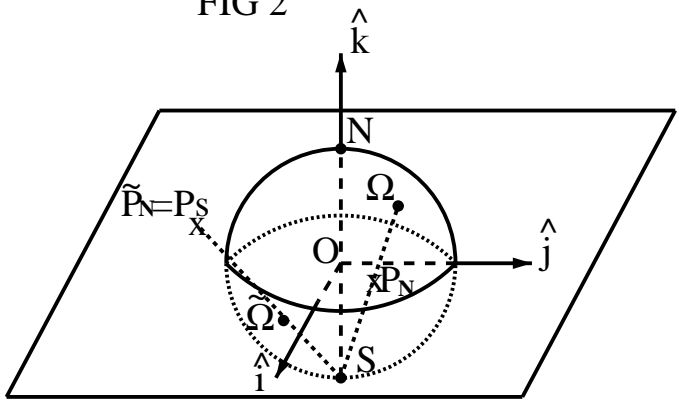
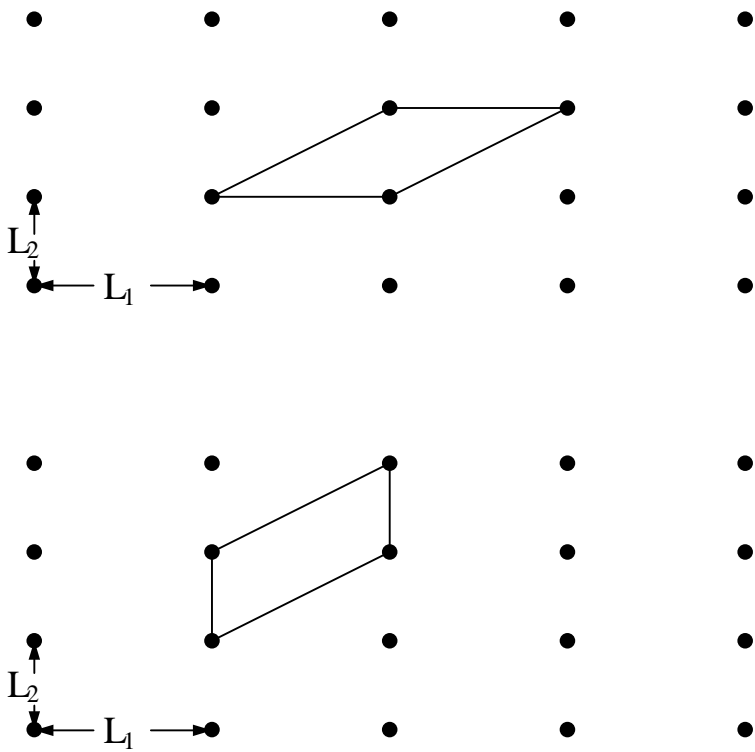
\end{document}